\documentclass[superscriptaddress,twocolumn,aps,pra,showpacs,
fixfloats]{revtex4-1}
\usepackage{amsmath}
\usepackage{bm}
\usepackage{dcolumn}
\usepackage{graphicx}

\begin{document}

\title{Featured trends in e$^{-}+$Mn electron elastic scattering}
\author{V. K. Dolmatov}
\affiliation{University of North Alabama,
Florence, Alabama 35632, USA}
\author{ M. Ya. Amusia}
\affiliation{Racah Institute of Physics, Hebrew University, 91904 Jerusalem, Israel}
\affiliation{A. F. Ioffe Physical-Technical Institute, 194021 St. Petersburg, Russia }
\author{L. V. Chernysheva}
\affiliation{A. F. Ioffe Physical-Technical Institute, 194021 St. Petersburg, Russia}

\date{\today}
\begin{abstract}
The impacts of both exchange interaction and electron correlation, as well as their combined impact,
on electron elastic scattering off a semifilled shell Mn(...$3d^{5}$$4s^{2}$, $^{6}S$)
 atom are theoretically studied in the electron energy range of $\epsilon= 0-25$ eV. Corresponding elastic scattering phase
 shifts $\delta_{\ell}(\epsilon)$ as well as partial $\sigma_{\ell}(\epsilon)$ and total $\sigma(\epsilon)$
 cross sections are found to be subject to a strong correlation impact.
 The latter is shown to be drastically different for oppositely
 spin-polarized scattered electrons, in some cases, thereby bringing  significant differences in  corresponding
 $\delta_{\ell}(\epsilon)$s,  $\sigma_{\ell}(\epsilon)$s, and $\sigma(\epsilon)$s between  said electrons. This is
 proven to be an inherent features of electron scattering off a semifilled shell atom in general. Electron correlation is accounted for
 in the framework of the self-energy part $\Sigma$ of the Green function of a scattered electron concept. The latter is calculated both in
 the second-order perturbation theory in the Coulomb interelectron interaction as well as beyond it by
 solving the Dyson equation for $\Sigma$.  The significance of the ``Dyson'' correlation corrections in e$^{-}+$Mn scattering
 is unraveled. They are shown
 to aggravate noticeably the inherent differences between elastic scattering phase shifts and cross sections of
spin-up ($\uparrow$) and spin-down ($\downarrow$) polarized electrons scattered off a spin-polarized Mn atom, in some cases.
In particular,
the existence of a narrow resonant maximum in $\sigma^{\downarrow}(\epsilon)$
near $\epsilon \approx 8$ eV but the absence of such in $\sigma^{\uparrow}(\epsilon)$ in e$^{-}+$Mn scattering is predicted.
\end{abstract}

\maketitle

\section{Introduction}

The $3d^{5}$ semifilled shell Mn(...$3d^{5}$$4s^{2}$, $^{6}S$) atom  has long served as the bridge
to, and touchstone for, a better understanding
of the interaction of transition metal atoms with X-ray and Vacuum-Ultraviolet
radiations from early days (see, e.g., works by Connerade et.\,al. \cite{Connerade}, Davis and Feldkamp \cite{Davis}, Amusia et.\,al. \cite{Amusia83})
to now (see review papers by Sonntag and Zimmermann \cite{Sonntag}, Martins et.\,al. \cite{Martins},
as well as some of the most recent papers by Frolov et.\,al. \cite{Manakov2010}, Osawa et.\,al. \cite{Osawa12},  Hirsch et.\,al. \cite{Hirsch2p'12} and references therein). The structure and
spectra of the Mn atom are of self-interest as well, in view of the found abundance of unique features associated with its semifilled $3d^{5}$ subshell.

In contrast, studies of another process of the basic and applied significance - electron scattering off the Mn atom - are too scarce. The Mn atom presents
a special interest for studying electron scattering processes. This is because it belongs to a class of atoms with the highest spin multiplicity. The latter is owing to co-directed spins of all five
electrons in the Mn $3d^{5}$ subshell, due to Hund's rule. As of today, understanding of electron scattering off   Mn   is rudimentary. The only available
experimental data relate to corresponding e$^{-}+$Mn differential scattering cross sections (DSC) measured at a \textit{single} value of the electron energy $\epsilon=20$ eV
 by Williams et.\,al. \cite{Williams} and Meintrup et.\,al. \cite{Meintrup}. The same stands for theoretical studies as well.
Thus, to understand and interpret experiment, Amusia and Dolmatov \cite{JETP90}
calculated the $20$ eV e$^{-}+$Mn elastic DCS  in the framework of a multielectron
simplified ``spin-polarized'' random phase approximation with exchange (SPRPAE) \cite{ATOM}, Meintrup et.\,al. \cite{Meintrup} did  it in a R-matrix framework,
and recently Remeta and Kelemen \cite{Remeta2010} performed their calculations of said DCS
in the framework of a local spin density approximation, but, again, only at discrete values of the electron energies of $10$ and $20$ eV.

 Whatever interesting effects were found in the above
 cited works they provide only a limited insight into a problem, since things might work quite differently at other, especially lower electron energies but just $10$ and $20$ eV energies.  Clearly,
 studying the scattering process through a continuum spectrum of electron energies is a way toward a deeper understanding of, as well as discovering new trends in, e$^{-}+$Mn elastic scattering,
 in particular, and e$^{-}+$$\textit{any semifilled shell atom}$, in general. However, no such study has been performed to date, to the  best of these authors' knowledge. Its performance is,
 thus, not merely a wish but necessity.

 It is the ultimate aim of the present paper to study e$^{-}+$Mn elastic scattering through a continuum spectrum of electron energies, to fill in an important gap in the
 current knowledge on said process. To meet this end, we focus on calculations of e$^{-}+$Mn elastic scattering phase shifts $\delta_{\ell}(\epsilon)$ and corresponding total cross section $\sigma(\epsilon)$ in the electron
 energy range of $0-25$ eV.

 The performed calculations, following the work \cite{ATOM},
 utilize a concept of the reducible self-energy part $\tilde{\Sigma}(\epsilon)$ of the Green function $G$ of an incoming electron. The calculations are carried out in three consequentially growing levels
 of sophistication. First, this is
 a one-electron ``spin-polarized'' Hartree-Fock (SPHF) approximation \cite{Slater} which is the zero-order approximation in perturbation theory in the Coulomb interelectron interaction $V$ for the Green function.
 In Refs.~\cite{JETP90,Remeta2010}, SPHF was adapted, as well as proven to be applicable, to the description of elastic electron scattering off semifilled shell atoms. Second, $\tilde{\Sigma}(\epsilon)$ is calculated in the second-order approximation
 in perturbation theory in $V$, to account for electron correlation in the system. This approximation is known as a simplified random phase
 approximation with exchange \cite{ATOM} and, with SPHF being chosen as the zero-order approximation with respect to
   electron correlation interactions, is referred to as SPRPAE$1$, in the present paper. Third, for a fuller account of electron correlation in electron scattering, $\tilde{\Sigma}(\epsilon)$ is determined
 beyond SPRPAE$1$ approximation  by solving the Dyson equation for $\tilde{\Sigma}(\epsilon)$ of a scattered electron, as in Ref.~\cite{ATOM}. Such approximation is referred to as SPRPAE$2$ in the paper. By
 comparing results obtained in the framework of each of the SPHF, SPRPAE$1$, and SPRPAE$2$ approximations we unravel several important trends in e$^{-}+$Mn scattering; these have already been noted in Abstract and are subject to a detailed discussion in text.

 Atomic units (a.u.) are used throughout the paper unless specified otherwise.

\section{Review of theory}

\subsection{SPHF}

A convenient starting approximation to study the structure and spectra of semifilled shell atoms is a ``spin-polarized'' Hartree-Fock (SPHF) approximation \cite{Slater}. Over the years, SPHF has been extensively and successfully exploited by the authors of this paper and their colleagues  (see, e.g., Refs.~\cite{Amusia83,Mn3s,Mn+,Cr,Cr+}) by using it directly, or  utilizing it as the zero-order approximation for a multielectron random phase approximation with exchange (RPAE) \cite{ATOM}, to study photoionization of said atoms and their ions. As noted earlier,
SPHF was also used successfully by Amusia and Dolmatov \cite{JETP90} to provide the initial understanding and interpretation of experimental data of Williams et.\,al. \cite{Williams}
on the $20$ eV e$^{-}+$Mn elastic DCS. Recently, SPHF has been employed by Remeta and Kelemen \cite{Remeta2010} to adapt a local density approximation to electron scattering off semifilled shell atoms, including Mn, to provide a further understanding of results of experiments \cite{Williams,Meintrup}. Applicability of SPHF, as well as the generalized on its basis
 other methods \cite{Amusia83,JETP90,Remeta2010}, for calculations and understanding of said phenomena with semifilled shell atoms-participants is out of doubt.

The quintessence of SPHF is as follows. It accounts for the fact that spins of all
electrons in a semifilled subshell of the atom (e.g., in the $3d^{5}$ subshell of Mn) are co-directed, in accordance with Hund's rule. It will be assumed in the present paper that spins of said electrons point upward ($\uparrow$). This results in splitting of each of other
closed ${n\ell}^{2(2\ell+1)}$ subshells in the atom into two semifilled subshells of opposite spin orientations, ${n\ell}^{2\ell+1}$$\uparrow$ and ${n\ell}^{2\ell+1}$$\downarrow$. This is in view of
 the presence of
exchange interaction between $nl$$\uparrow$ electrons with   spin-up electrons in the original semifilled
subshell of the atom but  absence of such for $nl$$\downarrow$ electrons. Thus, the atom of concern of this paper - the Mn atom - has  the following SPHF configuration:
 Mn(...${3\rm p}^{3}$$\uparrow$${3\rm p}^{3}$$\downarrow$${3\rm d}^{5}$$\uparrow$${4\rm s}^{1}$$\uparrow$$4s^{1}$$\downarrow$, $^{6}$S).
SPHF equations for the ground or excited states of a semifilled shell atom differ from ordinary HF equations for closed shell atoms (see, e.g. \cite{ATOM}) by accounting for exchange interaction only between electrons with the same spin orientation
($\uparrow$, $\uparrow$ or $\downarrow$, $\downarrow$).

By solving SPHF equations, one determines radial parts $P^{\uparrow}_{\epsilon \ell}(r)$
and  $P^{\downarrow}_{\epsilon\ell}(r)$ of the wavefunctions of spin-up and spin-down electrons in the ground, excited, or  scattering state of the atom.
For the continuum energy spectrum of scattered electrons, $P_{\epsilon \ell}^{\uparrow(\downarrow)}(r)$ have the well-known for a central field asymptotic behavior at large $r\gg 1$:
\begin{eqnarray}
P^{\uparrow (\downarrow)}_{\epsilon\ell}(r) \approx \frac{1}{\sqrt{\pi k}}\sin\left(k r -\frac{\pi\ell}{2}+\delta^{(0)\uparrow(\downarrow)}_{\ell}(\epsilon)\right).
\label{Eq1}
\end{eqnarray}
Here, $k$, $\ell$, $\epsilon$, and $\delta^{{\rm SPHF}\uparrow(\downarrow)}_{\ell}(\epsilon)$ are the momentum, orbital momentum, energy, and
the phase shift of a scattered electron, respectively. The total electron elastic scattering cross sections of spin-up ($\sigma^{\uparrow}$) and spin-down ($\sigma^{\downarrow}$)
electrons are determined as
 \begin{eqnarray}
 \sigma^{\uparrow(\downarrow)}(k)= \frac{4\pi}{k^2}\sum^{\infty}_{\ell=0}(2\ell+1)\sin^{2}\delta_{\ell}^{\rm SPHF\uparrow(\downarrow)}(k).
 \label{Eq2}
 \end{eqnarray}

\subsection{SPRPAE$1$}

\begin{figure}[tbp]
\includegraphics[width=8cm]{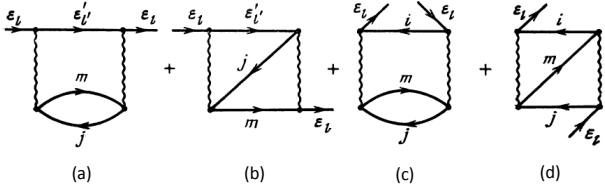}
\caption{The reducible self-energy part $\tilde{\Sigma}^{\rm SPRPAE1\uparrow(\downarrow)}(\epsilon)$ of the Green function of a scattered electron as defined in SPRPAE$1$. Here, a line with a right arrow denotes an electron, whether a scattered electron (lines
marked by $\epsilon_{l}$, $\epsilon'_{l'}$) or an atomic excited electron (a line $m$), a line with a left arrow denotes a vacancy (hole) in the atom
(lines $j$ and $i$), a wavy line denotes the Coulomb interelectron interaction $V$. On the other hand, the notations $\epsilon_{l}$, $\epsilon'_{l'}$, $m$, $j$, and $i$ themselves stand for corresponding
electronic states: $|\epsilon_{l}$$>$, $<$$j|$, \textit{etc}.}
\label{fig1}
\end{figure}

A simplified random phase approximation with exchange, version-$1$ (SPRPAE$1$), accounts for electron correlation in a e$^{-}+A$ system in the second-order perturbation theory in
the Coulomb interelectron interaction $V$ between the incoming and atomic electrons \cite{ATOM}. The approximation exploits the concept of
the reducible self-energy part of the one-electron Green function  $\tilde{\Sigma}(\epsilon)$ of a spin-up, $\tilde{\Sigma}^{\uparrow}(\epsilon)$,
 or spin-down, $\tilde{\Sigma}^{\downarrow}(\epsilon)$, scattered electron. In the framework of SPRPAE$1$, $\tilde{\Sigma}^{\uparrow(\downarrow)}(\epsilon)$ is illustrated
with the help of   Feynman  diagrams in Fig.~\ref{fig1}.

In Fig.~\ref{fig1}, diagrams (a) and (c) are called ``\textit{direct}'' diagrams, in contrast to ``\textit{exchange}'' diagrams (b) and (d). The latter two are due to exchange  interaction in a e$^{-}+A$ system.  Basically, the diagrams
in Fig.~\ref{fig1} illustrate how an incoming electron ``$\epsilon_{\ell}$'' polarizes a $j$-subshell in the atom by causing actual or virtual excitations $j$$\rightarrow$$m$ from the subshell, and couples with these excited states itself via
the Coulomb interaction. Note, exchange diagrams (b) and (d) in Fig.~\ref{fig1} vanish whenever spin directions of an incoming electron and polarized subshell of the atom are
opposite, due to orthogonality of electron spin-functions. We do not account for spin-flip effects since they are  negligible in e$^{-}+$Mn scattering \cite{Meintrup}, at the energies of interest.

Once $\tilde{\Sigma}^{\uparrow(\downarrow)}$ is calculated, the elastic electron scattering phase shifts of spin-up and spin-down
electrons are determined as \cite{ATOM}
\begin{eqnarray}
\delta_{\ell}^{\uparrow(\downarrow)} = \delta_{\ell}^{\rm SPHF\uparrow(\downarrow)} +\Delta\delta_{\ell}^{\uparrow(\downarrow)}.
\label{Eq3}
\end{eqnarray}
Here,  $\Delta\delta_{\ell}^{\uparrow(\downarrow)}$ is the correlation correction term to the SPHF calculated phase shift $\delta_{\ell}^{\rm SPHF\uparrow(\downarrow)}$:
\begin{eqnarray}
\Delta\delta_{\ell}^{\uparrow(\downarrow)} = \tan^{-1}\left(-\pi \left<\epsilon\ell^{\uparrow(\downarrow)}|\tilde{\Sigma}^{\uparrow(\downarrow)}|\epsilon\ell^{\uparrow(\downarrow)}\right>\right).
\label{Eq4}
\end{eqnarray}
A mathematical expression for the matrix element
 $~\left<\epsilon\ell^{\uparrow(\downarrow)}|\tilde{\Sigma}^{\uparrow(\downarrow)}|\epsilon\ell^{\uparrow(\downarrow)}\right >$
 is obtained with the help of the many-body correspondence rules \cite{ATOM}; we refer the reader  for details  to Ref.~\cite{ATOM}.

When the energy of a scattered electron exceeds the ionization threshold of the atom-scatterer, the term $\Delta\delta_{\ell}^{\uparrow(\downarrow)}$ and, thus,
the phase shift $\delta_{\ell}^{\uparrow(\downarrow)}$ itself become complex \cite{ATOM}.  Correspondingly,
\begin{eqnarray}
\delta_{\ell}^{\uparrow(\downarrow)} = \lambda_{\ell}^{\uparrow(\downarrow)} +i\mu_{\ell}^{\uparrow(\downarrow)}.
\label{Eq6}
\end{eqnarray}
Here,   $\lambda_{\ell}^{\uparrow(\downarrow)}$ and   $\mu_{\ell}^{\uparrow(\downarrow)}$ are the real and imaginary parts of $\delta_{\ell}^{\uparrow(\downarrow)}$, respectively:
\begin{eqnarray}
\lambda_{\ell}^{\uparrow(\downarrow)} = \delta_{\ell}^{\rm SPHF\uparrow(\downarrow)} + \Re\Delta\delta_{\ell}^{\uparrow(\downarrow)}, \quad \mu_{\ell}^{\uparrow(\downarrow)} = \Im \Delta\delta_{\ell}^{\uparrow(\downarrow)}.
\label{Eq7}
\end{eqnarray}
The total electron elastic scattering cross section is then determined \cite{ATOM} by
\begin{eqnarray}
\sigma = \frac{2\pi}{k^2}\sum_{\ell =0}^{\infty}(2\ell+1)(\cosh{2\mu_{\ell}}-
\cos{2\lambda_{\ell}}){\rm e}^{-2\mu_{\ell}}.
\label{Eq8}
\end{eqnarray}
In the context of the present paper, $\sigma \equiv \sigma^{\uparrow(\downarrow)}$, $\mu_{\ell} \equiv \mu_{\ell}^{\uparrow(\downarrow)}$, and $\lambda_{\ell} \equiv \mu_{\ell}^{\uparrow(\downarrow)}$.

\subsection{SPRPAE$2$}

A version-$2$ of the simplified random phase approximation with exchange (SPRPAE$2$) provides a fuller account for electron correlation. There, the reducible self-energy part of the one-electron Green function of a scattered electron
is sought as the solution of corresponding Dyson equation \cite{ATOM}. The latter, in an operator form, in terms of spin-up and spin-down electrons, is
\begin{eqnarray}
\hat{\tilde{\Sigma}}^{\uparrow(\downarrow)} = \hat{\Sigma}^{\uparrow(\downarrow)} - \hat{\Sigma}^{\uparrow(\downarrow)}\hat{G}^{\rm SPHF\uparrow(\downarrow)}\hat{\tilde{\Sigma}}^{\uparrow(\downarrow)}.
\label{Eq9}
\end{eqnarray}
Here, $\hat{\Sigma}^{\uparrow(\downarrow)}$ and  $\hat{\tilde{\Sigma}}^{\uparrow(\downarrow)}$  are the operators of the irreducible and reducible self-energy components of the Green-function operator for an incoming electron, respectively, and $\hat{G}^{\rm SPHF\uparrow(\downarrow)}$ is the operator of the Green function in the framework of SPHF:
$\hat{G}^{\rm SPHF\uparrow(\downarrow)}= (\hat{H}^{\rm SPHF\uparrow(\downarrow})-\epsilon)^{-1}$, where
$\hat{H}^{\rm SPHF\uparrow(\downarrow)}$ is the SPHF Hamiltonian operator of the system.

In SPRPAE$2$, to avoid tremendous calculation difficulties, the general Dyson equation, Eq.~(\ref{Eq9}), is simplified \cite{ATOM}.
This is achieved by replacing the operator of the irreducible self-energy
component of the Green function $\hat{\Sigma}^{\uparrow(\downarrow)}$  by  $\tilde{\Sigma}^{\rm SPRPAE1\uparrow(\downarrow)}$ (see Fig.~\ref{fig1}), to a good
approximation. As in SPRPAE$1$, corresponding SPRPAE$2$ phase shifts $\delta_{\ell}^{\uparrow(\downarrow)}$ and total cross sections
$\sigma^{\uparrow(\downarrow)}$ are determined by Eqs.~(\ref{Eq3})-(\ref{Eq8}).

\section{Results and discussion}

\subsection{Elastic scattering phase shifts}

\begin{figure}[tbp]
\includegraphics[width=7cm]{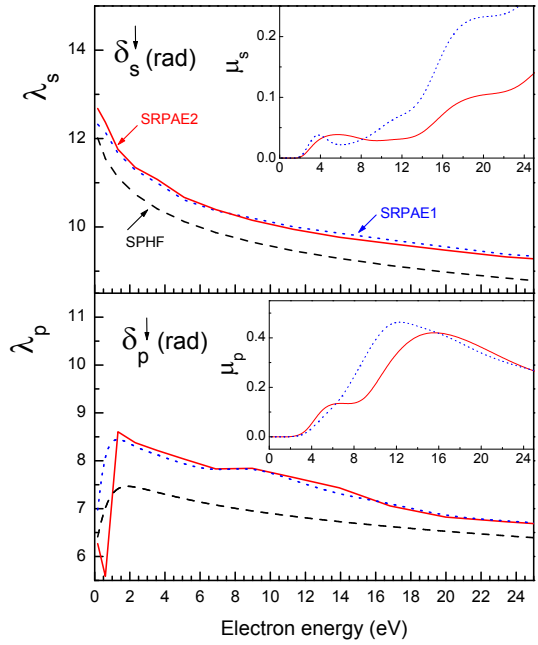}
\caption{(Color online)
SPHF (dashed line), SPRPAE$1$ (dotted line) and SPRPAE$2$ (solid line) calculated data for real ($\lambda_{\ell}^{\downarrow}$) and imaginary ($\mu_{\ell}^{\downarrow}$) parts of the e$^{-}+$Mn elastic scattering
phase shifts $\delta_{\ell}^{\downarrow}(\epsilon)=\lambda_{\ell}(\epsilon)+ i\mu_{\ell}(\epsilon)$ of $s$ and $p$ spin-down electronic waves versus the electron energy $\epsilon$.}
\label{fig2}
\end{figure}

\begin{figure}[tbp]
\includegraphics[width=7cm]{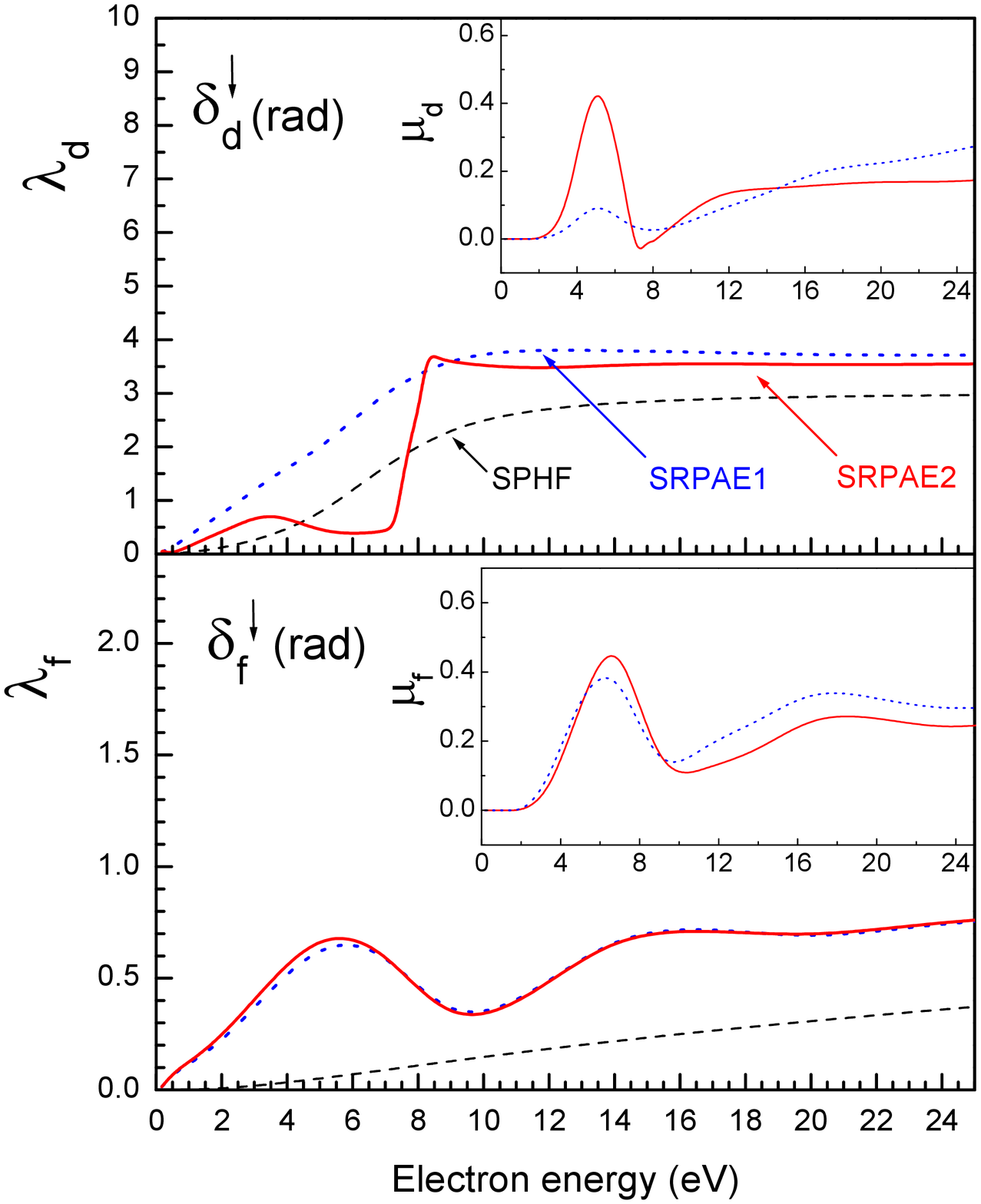}
\caption{(Color online)
SPHF (dashed line), SPRPAE$1$ (dotted line) and SPRPAE$2$ (solid line) calculated data for real ($\lambda_{\ell}^{\downarrow}$) and imaginary ($\mu_{\ell}^{\downarrow}$) parts of the e$^{-}+$Mn elastic scattering
phase shifts $\delta_{\ell}^{\downarrow}(\epsilon)$ of $d$ and $f$ spin-down electronic waves.}
\label{fig3}
\end{figure}

\begin{figure}[tbp]
\includegraphics[width=7cm]{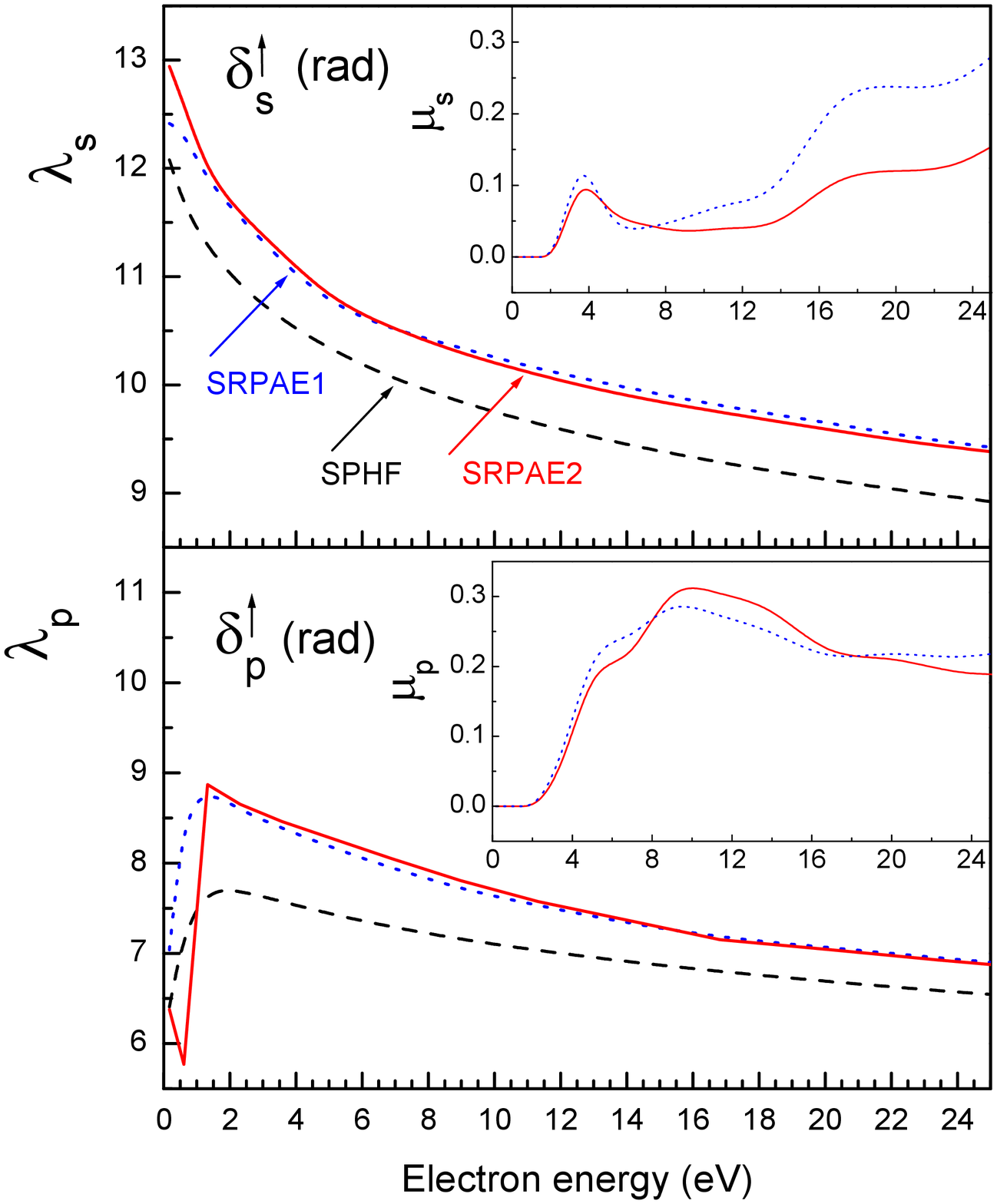}
\caption{(Color online)
SPHF (dashed line), SPRPAE$1$ (dotted line) and SPRPAE$2$ (solid line) calculated data for real ($\lambda_{\ell}^{\uparrow}$) and imaginary ($\mu_{\ell}^{\uparrow}$) parts of the e$^{-}+$Mn elastic scattering
phase shifts $\delta_{\ell}^{\uparrow}(\epsilon)$ of $s$ and $p$ spin-up electronic waves.}
\label{fig4}
\end{figure}

\begin{figure}[tbp]
\includegraphics[width=7cm]{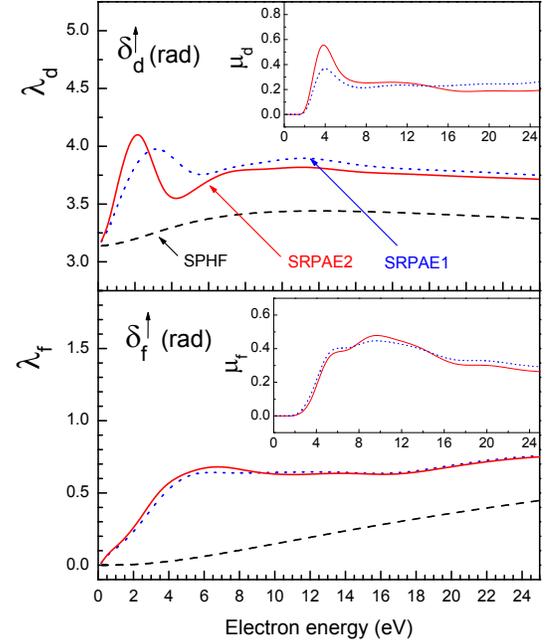}
\caption{(Color online)
SPHF (dashed line), SPRPAE$1$ (dotted line) and SPRPAE$2$ (solid line) calculated data for real ($\lambda_{\ell}^{\uparrow}$) and imaginary ($\mu_{\ell}^{\uparrow}$) parts of the e$^{-}+$Mn elastic scattering
phase shifts $\delta_{\ell}^{\uparrow}(\epsilon)$ of $d$ and $f$ spin-up electronic waves.}
\label{fig5}
\end{figure}

In the present paper, e$^{-}+$Mn elastic electron scattering is investigated in the electron energy range between approximately $0$ and $25$ eV, as a case study. This is because core-polarization effects
in e$^{-}+$Mn elastic electron scattering are expected to be particularly strong at low electron energies. A trial calculation showed that, at the given energies,
accounting for contributions of only $s$, $p$, $d$, and $f$ partial waves to the total elastic scattering cross section, as well as accounting for only monopole, dipole,
 quadrupole, and octupole excitations of the atomic core in calculations of $\hat{\tilde{\Sigma}}^{\uparrow(\downarrow)}$ is an excellent approximation.
SPHF, SPRPAE$1$, and SPRPAE$2$ calculated data for real ($\lambda_{\ell}^{\downarrow}$) and imaginary ($\mu_{\ell}^{\downarrow}$) parts of elastic scattering
phase shifts $\delta_{\ell}^{\downarrow}(\epsilon)$ of spin-down electronic waves are depicted in Figs.~\ref{fig2} and \ref{fig3}, and those for
$\lambda_{\ell}^{\uparrow}$ and $\mu_{\ell}^{\uparrow}$ of $\delta_{\ell}^{\uparrow}$ for spin-up electronic waves - in Figs.~\ref{fig4} and \ref{fig5}.


The depicted   data unravel  important trends in low energy e$^{-}+$Mn elastic scattering which are detailed below.

\subsubsection{The effects of electron correlation in e$^{-}+$Mn elastic scattering}

First, one can see from Figs.~\ref{fig2}-\ref{fig5} that both SPRPAE$1$ and SPRPAE$2$ correlation
 affects strongly all phase shifts quantitatively and in a number of cases - for $p$, $d$, and $f$ partial waves - even qualitatively, compared to SPHF calculated data.
 Thus, the utter importance of electron correlation in the e$^{-}+$Mn low energy elastic electron scattering is revealed.

Second,  notice how SPRPAE$2$ calculated data for low energy $\delta^{\uparrow\downarrow}_{d}(\epsilon)$  phase shifts differ drastically  from calculated data
obtained in the framework of SPRPAE$1$. Indeed, the SPRPAE$2$ calculated phase shift $\delta^{\downarrow}_{d}(\epsilon)$ (Fig.~\ref{fig3}) drops abruptly,  with decreasing electron energy,  from
$\delta^{\downarrow}_{d}(\epsilon) \approx 3.5$  to $\delta^{\downarrow}_{d}(\epsilon) \approx 0.5$ rad between $8$ to $7$ eV, then
 develops a   maximum at yet lower energies, after which it approaches its final value of $\delta^{\downarrow}_{d}(\epsilon) \approx 0$ at $\epsilon=0$. Clearly, the described behavior of
SPRPAE$2$ calculated $\delta^{\downarrow}_{d}(\epsilon)$ has little in common with that obtained in the framework of SPRPAE$1$. Strong quantitative differences take place between SPRPAE$1$ and SRPAR$2$
calculated  data for $\delta^{\uparrow}_{d}(\epsilon)$ as well, see Fig.~\ref{fig5}.
Also, notice  abrupt  changes in SPRPAE$2$ calculated phase shifts $\delta^{\uparrow\downarrow}_{p}(\epsilon)$
between approximately $0$ and $2$ eV, in opposition to those calculated in SPRPAE$1$. The unraveled drastic differences between SPRPAE$2$ and SPRPAE$1$ calculated   phase shifts
is a novel result; it was not observed in previous similar calculations performed for other atoms. Thus, the present work reveals, and generally proves, the necessity for a fuller account (as in SPRPAE$2$) of correlation beyond the second-order
 approximation (SPRPAE$1$) for an adequate understanding of e$^{-}+$Mn  scattering.

Third, which is  of significant importance, notice differences between phases of scattered electronic waves with the same $\ell$s but opposite spin orientations, $\delta_{\ell}^{\downarrow}$ versus
$\delta_{\ell}^{\uparrow}$.
 Said differences are drastic for $\epsilon d$ as well as $\epsilon f$ spin-up and spin-down waves
 in all three SPHF, SPRPAE$1$, and SPRPAE$2$ approximations (cp. Figs.~\ref{fig3} and \ref{fig5}). For other spin-up and spin-down waves the differences in question are less spectacular, but, nevertheless,
 exist as well. Thus, it is revealed in the present study that, generally, scattering of oppositely spin-polarized electrons off the Mn atom take  different routes.

 Below, we provide reasons for the unraveled trends in the e$^{-}+$Mn electron scattering phase shifts.

\subsubsection{The origin of the zero-order ({\rm SPHF}) difference between scattering of spin-up and spin-down electrons off Mn}

In SPHF, the dependence of e$^{-}+$Mn scattering phase shifts on a spin-orientation of a scattered electron is a straightforward effect. It is due to the presence/absence of exchange interaction between, respectively,
 $\epsilon\ell$$\uparrow$/$\epsilon\ell$$\downarrow$ incoming electrons and primarily five $3d$$\uparrow$ electrons in the semifilled $3d^{5}$$\uparrow$ subshell in the atom. This is  a
 characteristic feature of elastic electron scattering off any semifilled shell atom, for an obvious reason.

 Next, since no stable negative Mn ion exists with the ground-state atomic
core configuration \cite{Radtsig}, the number $q_{d\uparrow }$ of bound $d$$\uparrow$ states in the atom is $q_{d\uparrow }=1$ (due to the presence of $3d^{5}$$\uparrow $ subshell in the atom), whereas corresponding $q_{d\downarrow }=0$. Consequently, in accordance with the generalized
Levinson's theorem [$\delta _{\ell }(\epsilon )\rightarrow (n_{\ell}+q_{\ell })\pi $ as $\epsilon \rightarrow 0$, $n_{\ell }$ being the number
of bound states with given $\ell$ in the field of an atom and $q_{\ell }$
being the number of occupied $\ell $ states in the atom itself] \cite{Landau},
 $\delta _{d}^{\rm{SPHF\uparrow }}(\epsilon )\rightarrow \pi $, whereas
$\delta _{d}^{\rm{SPHF\downarrow }}(\epsilon )\rightarrow 0$ as $\epsilon \rightarrow 0$. This translates into the initial drastic
quantitative and qualitative differences between SPHF calculated phase
shifts $\delta _{d}^{\rm{SPHF\downarrow }}(\epsilon )$ (Fig.~\ref{fig3},
dashed line) and $\delta _{d}^{\rm{SPHF\uparrow }}(\epsilon )$ (Fig.~\ref{fig5}, dashed line).

\subsubsection{The origin of the second-order ({\rm SPRPAE$1$}) difference between scattering of spin-up and spin-down electrons off Mn}

In SPRAE$1$, compared to SPHF, differences between $\delta_{d}^{\downarrow}$ and $\delta_{d}^{\uparrow}$ scattering phases become even more spectacular.  Indeed, one can see (Fig.~\ref{fig3}, dotted line) that the
SPRPAE$1$ calculated real part $\lambda_{d}^{\downarrow}$ of $\delta_{d}^{\downarrow}$ is a monotonic function of $\epsilon$ in the interval of
$8$ to $0$ eV, but the real part $\lambda_{d}^{\uparrow}$ of $\delta_{d}^{\uparrow}$ (Fig.~\ref{fig5}, dotted line) is not. The latter now has a well developed minimum followed by an appreciable maximum with decreasing energy $8 \rightarrow 0$ eV. This results, partly, in an additional, compared to SPHF, quantitative as well as qualitative difference between  $\delta_{d}^{\downarrow}$ and $\delta_{d}^{\uparrow}$, due to
the effects of electron correlation.

Next, when SPRPAE$1$ correlation is accounted for, no lesser spectacular differences also emerge between the calculated $\delta_{f}^{\downarrow}$ and $\delta_{f}^{\uparrow}$ scattering phases.
Indeed, SPRPAE$1$ correlation
is seen to induce strong differences between the real parts $\lambda_{f}^{\downarrow}$ (Fig.~\ref{fig3}, dotted line) and $\lambda_{f}^{\uparrow}$
(Fig.~\ref{fig5}, dotted line) of phase shifts  $\delta_{f}^{\downarrow}$ and $\delta_{f}^{\uparrow}$: $\lambda_{f}^{\rm SPRPAE1\downarrow}$ turns into an oscillating function of $\epsilon$
whereas $\lambda_{f}^{\rm SPRPAE1\uparrow}$ does not.

A trial calculation showed that the SPRPAE$1$ correlation induced differences between $\lambda_{\ell}^{\rm SPRPAE1\downarrow}$ and $\lambda_{\ell}^{\rm SPRPAE1\uparrow}$
are due primarily to polarization of the $3d^{5}$$\uparrow$ subshell by an incoming $\epsilon\ell$ electron. Thus, when the incoming electron is a spin-up electron,
exchange diagrams (b) and (d) with $j=\left|3d\right.$$\uparrow$$\left. \right>$ in Fig.~\ref{fig1} contribute to phase shifts, whereas their contributions vanish for a spin-down incoming electron, as was explained earlier in text.

\subsubsection{The higher-order ({\rm SPRPAE$2$}) correlation difference between scattering of spin-up and spin-down electrons off Mn}

Higher order SPRPAE$2$ correlation corrections, compared to SPRPAE$1$, induce additional significant changes, primarily in $\delta_d$$^{\downarrow}$ and $\delta_d$$^{\uparrow}$   scattering
phases.
While the impact of said corrections on a real part $\lambda_d$$^{\uparrow}(\epsilon)$ of $\delta_d$$^{\uparrow}$ (Fig.~\ref{fig5}, solid line) results mostly in its quantitative change
compared to corresponding SPRPAE$1$ data,
it (the impact) makes a real part $\lambda_d$$^{\downarrow}(\epsilon)$ of $\delta_d$$^{\downarrow}$ to become both quantitatively and qualitatively different than
$\lambda_d$$^{\rm SPRPAE1\downarrow}$, Fig.~\ref{fig3}, solid line.

The unraveled different impact of SPRPAE$2$ correlation on $\delta_d$$^{\downarrow}$ compared to $\delta_d$$^{\uparrow}$ is because the SPRPAE$2$ equation for the reducible self-energy part of the Green-function,
Eq.~(\ref{Eq9}), contains many cross-product terms between terms associated with each of the diagrams depicted in Fig.~\ref{fig1}. Note, there are no cross-products terms in the framework
of SPRPAE$1$ at all. Since the number of $3d$ electrons in the Mn atom is a $100\%$ unbalanced in favor
of $3d$$\uparrow$ electrons, many SPRPAE$2$ cross-product terms, involving various excitations of $3d$$\uparrow$ electrons, vanish for an incoming spin-down electron but remain for a spin-up scattered electron. This
aggravates the difference between the SPRPAE$2$ correlation impact on scattering of spin-up compared to spin-down electrons off the atom compared to the difference emerging in SPRPAE$1$ calculations. For $\epsilon d$ incoming electrons, this
difference is huge, as was illustrated above.

 Thus, the carried out study reveals that the lower-order SPRPAE$1$ approximation is clearly insufficient for a proper understanding of electron scattering off a semifilled shell atom. This is an
 important finding.

Note, the demonstrated fast variation of the SPRPAE$2$ $\delta_{d}^{\downarrow}$ phase with energy in a quite
narrow energy region near $8$ eV can be interpreted as a prominent time-delay of the
partial $\epsilon d$$\downarrow$ wave while crossing the atomic region. Indeed, long ago,
 Wigner \cite{Wigner} connected the time of quantum mechanical reaction
procedure $\tau _{r}$ with the derivative on the phase over energy. In the
considered case of up and down electrons, one has
\begin{eqnarray}
\tau _{r\ell}^{\uparrow(\downarrow)}=\frac{d\delta_{\ell}^{\uparrow(\downarrow)}}{d\epsilon}
\end{eqnarray}
With the help of this expression, one can estimate not only the time duration of the
scattering process of a given $\epsilon\ell$ wave, but also the time duration difference $\Delta\tau_{\ell}^{\uparrow\downarrow}$ for up and
down partial waves $\ell$. For the $\epsilon d$$\uparrow$ and $\epsilon d$$\downarrow$ waves in $e^{-}+$Mn scattering near $\epsilon \approx 8$ eV, we find that $\Delta_{d}^{\uparrow\downarrow} \approx 2$\, $10^{-15}$ s,
in the framework of SPRPAE$2$.

\subsection{Total e$^{-}+$Mn elastic scattering cross section}

\begin{figure}[tbp]
\includegraphics[width=7cm]{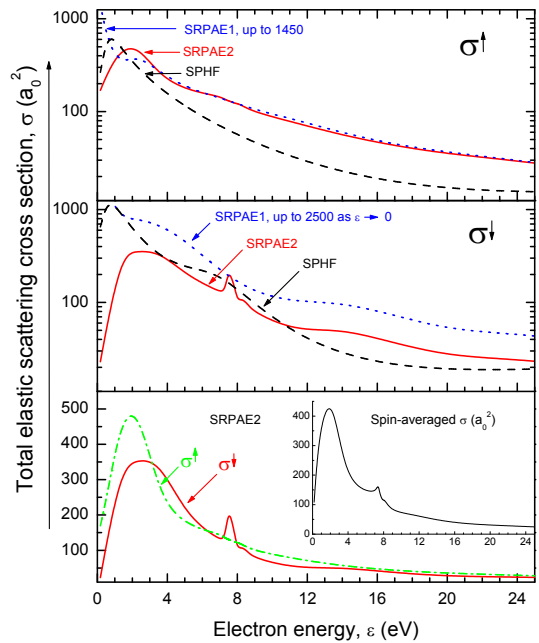}
\caption{(Color online)
Top panel: SPHF (dashed line), SPRPAE$1$  (dotted line) and SPRPAE$2$ (solid line) calculated data for the total elastic scattering cross section $\sigma^{\uparrow}(\epsilon)$ of
  spin-up electrons off the Mn atom. Middle panel: the same as in upper panel but for $\sigma^{\downarrow}(\epsilon)$ of spin-down electrons. Bottom panel: Individual data for
$\sigma^{\uparrow}(\epsilon)$ and $\sigma^{\downarrow}(\epsilon)$ calculated in the framework of SPRPAE$2$ (for a clearer comparison of these two final results). Inset: the total spin-averaged
elastic scattering cross section $\sigma^{\rm avrg}(\epsilon)$, see Eq.~(\ref{Eq10}), calculated in the framework of SPRPAE$2$.}
\label{fig6}
\end{figure}

To get insight into both the individual correlation and exchange interaction impacts, as well as their combined effect, on \textit{observable} elastic electron  scattering characteristics, we calculated
the e$^{-}+$Mn total spin-up $\sigma^{\uparrow}$ and spin-down $\sigma^{\downarrow}$ elastic scattering cross sections in the framework of  SPHF, SPRPAE$1$, and SPRPAE$2$.
The performed calculations utilized the above presented data for phase shifts. The thus calculated $\sigma^{\uparrow}$ and $\sigma^{\downarrow}$ are depicted in Fig.~\ref{fig6}.

The depicted results are  self-illustrating in the demonstration of the found significance of correlation impacts on e$^{-}+$Mn elastic electron scattering cross sections of spin-up
and spin-down electrons. Clearly,
a fuller account of correlation, secured by the SPRPAE$2$ approximation, is seen to be decisive.

Of significant interest is unraveling of the existence of a narrow resonance in $\sigma^{\downarrow}$ near $\epsilon = 8$ eV in the framework
of SPRPAE$2$ (see the middle and bottom panels in Fig.~\ref{fig6}). This resonance has an interesting nature which is   associated both with a semifilled shell structure of,
and electron correlation in, the Mn atom.

Indeed, as was discussed and demonstrated above, Fig.~\ref{fig3}, it is because of the semifilled shell structure of the Mn atom that the elastic scattering phase shift $\delta_{d}^{\rm SPHF\downarrow}$
as well as real parts $\lambda_{d}^{\downarrow}$ of both $\delta_{d}^{\rm SPRPAE1\downarrow}$ and  $\delta_{d}^{\rm SPRPAE2\downarrow}$ phase shifts drop to a zero at $\epsilon = 0$. On the way to a zero,
 $\delta_{d}^{\downarrow}$ as well as  $\lambda_{d}^{\downarrow}$ pass through the value of $\delta_{d}^{\downarrow}(\lambda_{d}^{\downarrow})=\pi/2$, at certain values of
 the electron energy. Correspondingly, at such energy, a resonance emerges in a partial $\sigma_{d}^{\downarrow}$ elastic scattering cross section both in SPHF,
 where $\sigma_{d}^{\downarrow} \propto \sin^{2}\delta_{d}^{\downarrow}$, Eq.~(\ref{Eq2}), and SPRPAE$1$ or SPRPAE$2$, where $\sigma_{d}^{\downarrow}$ depends on $\cos{2\lambda_{d}^{\downarrow}}$, Eq.~(\ref{Eq8}). Naturally,
 the existence of this resonance in partial $\sigma_{d}^{\downarrow}$ translates into its emergence in the total elastic scattering cross $\sigma^{\downarrow}$ section at the same
 electron energy as well. Indeed, as seen in Fig.~\ref{fig6}, the resonance in question is present in $\sigma^{\rm SPHF\downarrow}$ and $\sigma^{\rm SPRPAE2\downarrow}$ at $\epsilon \approx 8$ eV, as well
  as in $\sigma^{\rm SPRPAE1\downarrow}$ at $\epsilon \approx 5$ eV.
  The resonance is broad and weakly  developed both in the SPHF and SPRPAE$1$ calculated cross sections. In contrast, it is
 sharp and well developed in  $\sigma^{\rm SPRPAE2\downarrow}$. SPRPAE$2$, without doubt, is a more complete approximation than either of the SPHF or SPRPAE$1$ approximations. Prediction based on the basis of SPRPAE$2$
  should, thus, match reality better than those in the framework of SPHF or SPRPAE$1$. The SPRPAE$2$ calculated results in question, thus, allow us to claim the discovery of the \textit{actual} existence of the $8$ eV \textit{sharp} resonance in the e$^{-}+$Mn total elastic electron scattering cross section.
  Furthermore, as follows from the
 discussion, the resonance in question can exist neither without the semifilled shell nature of the Mn atom nor without accounting for electron correlation in the e$^{-}+$Mn scattering
 process. This permits us to speak about the discovery of a novel type of a resonance in electron-atom scattering which we name the \textit{semifilled-shell-correlation} resonance, to
 stress its uniqueness.

Note, the semifilled-shell-correlation resonance shows up not only in the total elastic electron scattering cross section $\sigma^{\downarrow}$ of spin-down electrons but in corresponding spin-averaged
total cross section $\sigma^{\rm avrg}$ as well, see inset in
bottom panel of Fig.~\ref{fig6}. The latter - $\sigma^{\rm avrg}$ - was calculated as
\begin{eqnarray}
\sigma^{\rm avrg}(\epsilon) = A_{S_{A}+s}\sigma^{\uparrow}(\epsilon) + A_{S_{A}-s}\sigma^{\downarrow}(\epsilon),
\label{Eq10}
\end{eqnarray}
where
\begin{eqnarray}
A_{S_{A} \pm s} = \frac{2 (S_{A}  \pm s)+1}{(2S_{A}+1)(2s+1)},
\label{Eq11}
\end{eqnarray}
$S_{A}$ being the spin of the atom ($S_{A}= 5/2$ for the Mn atom) and $s=1/2$ being the electron spin.

\section{Conclusion}

In summary, it is unraveled in the present paper that (a) SPRPAE$1$ and SPRPAE$2$ electron correlation affects significantly e$^{-}+$Mn electron elastic scattering phase shifts and cross sections,
(b) correlation may affect, and in the case of the Mn atom does affect, drastically differently scattering phase shifts of electrons with opposite spin-orientations, (c) a fuller account of correlation (as in SPRPAE$2$) beyond the second-order approximation of perturbation theory in the  Coulomb interelectron interaction (SPRPAE$1$) may be, and in the case of e$^{-}+$Mn scattering is, crucial for an adequate understanding of the spectrum of
corresponding  phase shifts and cross sections versus the electron energy, and (d) a combined impact of the semifilled shell nature of the Mn atom and electron correlation results in the emergence
of a novel type of the resonance - the semifilled-shell-correlation resonance - in the e$^{-}+$Mn total elastic scattering cross section.  We urge experimentalists and other theorists to verify the made predictions.

Furthermore, the unraveled physics behind the discovered  effects  in e$^{-}+$Mn scattering is so clear that the present authors have little doubts that most of the effects are inherent features
of electron elastic scattering off any multilectron semifilled shell atom in general, not just off the Mn atom. The predicted effects should even be stronger for atoms-scatterers with a higher spin multiplicity (than that of Mn), such as
 the Cr(...${3d}^{5}$$\uparrow$${4s}^{1}$$\uparrow$, $^{7}$S) or  Eu(...${4f}^{7}$$\uparrow$${6s}^{1}$$\uparrow$$6s^{1}$$\downarrow$, $^{8}$S) atoms, or similar. We are currently undertaking such study.

\section{Acknowledgements}
V.K.D. acknowledges the support of NSF Grants no.\ PHY-$0969386$ and PHY-$1305085$.

\end{document}